# Beyond Patient Monitoring: Conversational Agents Role in Telemedicine & Healthcare Support For Home-Living Elderly Individuals

Ahmed Fadhil
University of Trento Trento, Italy
fadhil@fbk.eu

*Abstract*—There is a need for systems to dynamically interact with ageing populations to gather information, monitor health condition and provide support, especially after hospital discharge or at-home settings. Several smart devices have been delivered by digital health, bundled with telemedicine systems, smartphone and other digital services. While such solutions offer personalised data and suggestions, the real disruptive step comes from the interaction of new digital ecosystem, represented by chatbots. Chatbots will play a leading role by embodying the function of a virtual assistant and bridging the gap between patients and clinicians. Powered by AI and machine learning algorithms, chatbots are forecasted to save healthcare costs when used in place of a human or assist them as a preliminary step of helping to assess a condition and providing self-care recommendations. This paper describes integrating chatbots into telemedicine systems intended for elderly patient after their hospital discharge. The paper discusses possible ways to utilise chatbots to assist healthcare providers and support patients with their condition.

## I. INTRODUCTION

The number of people need health support during or after hospitalisation has been considerably growing in the last decades. This has high care cost associated and overwhelms the healthcare service providers. People in need of care after discharge can range from kids, adolescents, adults and elderly. Regardless of age group, patients discharged always need healthcare support even after their treatment period. This support could be medical, but often its related to medication adherence and lifestyle promotion. Regular activities, such as meals, medication or social events, as well as extraordinary events like doctor's appointments are examples of support after hospital discharge.

Current telemedicine application are focused on ambient assistant living and supporting patients at home. However, enabling patients to stay autonomous in their home environment and to have a self-determined way of living for as long as possible is an important societal goal of the future healthcare. Several limitations are associated with telemedicine support. For example, providing an environment for health expert and patient is needed to interact with after hospital discharge. In addition, it's hard and inefficient to provide health support in rural area or areas with commuting difficulty.

Chatbot application can provide various support to patients living in rural areas, at home and facilitate tasks for the healthcare professionals. Integrating chatbot as a tool for elderly patients using telemedicine applications have the advantage of improved diagnosis and better treatment management, continuing education and training, quick and timely follow-up of discharged patients, and for healthcare professionals to access comprehensive patient data. Patients can benefit from chatbot systems to access to specialised healthcare services in underserved rural, semi-urban and remote areas, early diagnosis and treatment in case of deterioration, access to expertise of medical specialists, reduced physician's cost of visits, reduced visits to hospitals for minor cases, reduced travel expenses, early detection of disease, and provide a low cost health services [1]. The idea of integrating conversational agents in telemedicine systems can also help elderly patients who may be confused with a complex mHealth application UI, but certainly know how to write messages to their children or caregivers. Therefore, conversing with a health agent about their health concerns feels more natural to them. The problem with the personal relationship between a patient and their physician is that it is generally conducted in an 8-10 minute consultation, which is not a very satisfactory relationship. Conversational agent is perhaps what the patient with a life-threatening illness requires, when their insomnia at 2am caused by chemotherapy has them in despair. The chatbot can lend a sympathetic ear and record the problem, and suggest mitigations based on the patient's drug regime. Managing a patient's health is a 24/7 occupation, and the combination of chatbots, smart drugs and wearables can monitor the patient's condition and alert the physician to an impending medical crisis.

If done well, a healthcare provider can see upwards of up to 50% of traditional office visits converted to a less expansive, more convenient virtual visits [2]. Telemedicine can solve these pain points by bringing care to patients wherever they are and whenever they need it, while smoothing out the logistics of scheduling and traveling, so doctors can focus on their top priority of delivering care. Similarly, health AI holds promise of increasing efficiency in the care process for improved care outcomes and better time management.

With chatbots we are on the brink of a next paradigm shift, from the design of graphical layouts to AI-powered software where a conversation is the medium and content of user experiences. Therefore, chatbots as conversational agents are leveraging the potential of modern technologies to better mirror human capabilities [2], significantly lowering the entry barriers.

This paper describes current research in integrating chatbot application in telemedicine systems and addresses some areas where these can be applied to health care assistance for both at home and hospitalised patient care. We will highlight the benefit of such system to patients after hospital discharge and will discuss the current technology, behaviour change theories and linguistic requirements to build such architecture and integrate it into the healthcare services. In addition, we will discuss sentiment analysis from patient conversation and the role of human actor represented by healthcare provider in providing emotional support. This is because humans are not machines, and providing health support is complicated. Many psychological factors are influencing diseases, and hence patients need human contact in the loop. This work is a preliminary attempt at what we believe to be a large application area that can be applied towards assisted healthcare for elderly and people in rural areas.

## II. Telemedicine and Conversational Agents

A substantial body of work have been on special requirements for patients and people with impairments. A growing number of conversational AI have been applied in various domains, including education, business, entertainment and health [13], [14]. The concept of AI-based chatbot system is novel and still growing. However, chatbots can provide huge support to both healthcare organisations and patients. Chatbot systems are applied towards assisted healthcare and they facilitate doctors task and provide valuable health support to patients. For example, people with cognitive impairments can benefit from a chatbot as a daily assistant. However, we should consider the issue of user acceptance of the system to interact successfully. A work by Yaghoubzadeh et al., [3] performed interview with elderly and cognitively impaired users. The result from the interview and focus group showed that acceptance can be increased by ways of a participatory design method, which could be a good strategy to mitigate understanding problems. Several studies have investigated the benefit of virtual agents in providing services to users. The GUIDE project [4], also using focus groups, concluded spoken language was by far the most preferred interaction modality for elderly users unfamiliar with technology. Applications should dynamically interact with ageing populations to gather information, monitor their health condition and provide support whenever relevant. For those with chronic health issues or life-threatening illnesses, the line of specialists and procedures can seem endless. That said, each time a new practitioner or clinic is visited, the patient is required to fill in long and complicated forms. In order to cope, the industry has to undergo some big changes, starting with data management. Over the last decade, hospitals, medical practices, medical schemes and research facilities have been digitising their data, in an effort to warehouse it effectively. Apart from traditional patient data contained in text, there are various images and sounds recorded, from x-rays and ultrasound to MRI imaging. Some doctors even prefer that their conversations with patients be recorded for the patient's benefit. This collection of disparate information is generally unstructured and cannot be ordered in a neat table. Where healthcare entities have tackled this hurdle, the result is a holistic view of the patient, which removes some of the complexity of diagnosis for the medical practitioners and makes life simpler for the patient. It also opens the way for the move to machine-to-machine (M2M) communication and the use of artificial intelligence to sift through and analyse data transmitted from the sensors gathering it. The use of sensors to monitor everything from whether a patient took the right dose of medication at the right time to insulin levels is one of the big growing areas in the healthcare industry. Combining this with the global adoption of mobile devices, especially smartphones and wearables, we are moving to the time where our health can be monitored on a continuous and proactive basis by artificial intelligence (AI). Where a problem is detected, the healthcare professional can be alerted to take appropriate action, or a mitigation could even be instigated via AI, where appropriate. Considering elderly people as potential chatbot system users can offer a technology that enables users to overcome the confusing interfaces found in current human-computer interactions. Chatbots and virtual agents are increasingly utilised in training medical practitioners. A work by Kenny et al., [5] described virtual human technology developed for virtual patients for mental health diagnosis and clinician training. The paper provided insights on possible ways to utilise such systems for assisted healthcare.

The AI-powered chatbots have speech recognition, natural language and vision, machine learning, even sentiment analysis capabilities that allows them to interact with human users in a more natural way. Additionally, sensors are used to monitor the environment for specific behaviours that can be used into the chatbot system. As a result, the bot may respond to a patient or an elderly with a manner that assures powerful affect on their living situation. A work by Raij et al., [6] conducted two separate studies that compare interactions with a virtual human and with a real human in a medical interview scenario. Medical students interacted with either a virtual human simulating appendicitis or a real human pretending to have the same symptoms. The result showed similarity in content of the virtual and real interactions. However, participants appeared less engaged and insincere with the virtual human. There has also been considerable work on building socially interactive chatbots that exhibit human like qualities or to help patients deal with disabilities [8].

**Discussion**

Related works showed that including a chatbot in the patient-doctor interaction through a telemedicine application to support elderly after hospital discharge is a sound approach and can be a suitable interaction paradigm for the elderly. However, very little work has been on telemedicine systems for elderly after hospital discharge or who are live-in in rural areas. It seems reasonable to assume conversational AI can provide a suitable way of interacting with technology for those people too, who are often with technical impairments and can only use simple icons to interact with their technical environment. Chatbots have many advantages over existing mobile and web applications. They are easy to learn since their learning curve is low. Therefore, elderly users would accept it as an assistant to help with scheduling daily activities, checking with doctors, booking appointments, following medication, diet and physical activity. Since misunderstandings by the system are inevitable, especially with these user groups and when it comes to unrestricted spoken input in natural environments, a key sub-question is how such users react to misunderstandings of the chatbot system and whether specific interaction strategies of the agent can help them to spot and repair them in a way suited to their cognitive limitations. We shed light on these questions in the following sections.

## III. RESEARCH STUDY

### A. Why Conversational Agents

Healthcare sector is among the most overwhelmed by the number of patients who need continues social and emotional support outside the hospital settings. Unlike mobile apps, chatbots are easy to install and can deliver unique user experience. Moreover, two of the three apps on our phone are messaging apps. Using conversation is easier than heavy interfaces in the application especially by the elderly [9]. Using Conversational User Interfaces inside messaging apps prevents users from opening and closing a bunch of apps to get tasks accomplished. It allows us to invoke a text-message or simple button click and achieve the same task. People, including elderly spend more time on their chatting applications, such as WhatsApp, Telegram and Facebook than any other application. We conducted a questioner with 44 smartphone users, 30 were above 40 years old, and the rest were above 25. We asked them about the type of application they use the most (e.g., Social Networking (e.g., Telegram, WhatsApp), Medical (e.g., Diabetes apps), Entertainment (e.g., YouTube), Health and Fitness (e.g., Google Fit) and Games (e.g., any games)). The findings proved that chatting applications that also support conversational interfaces are widely used and people feel more comfortable using them on daily bases (see Table-I). The majority responded they use mostly social networking applications in their smartphone.

TABLE I. MOSTLY USED APPLICATION CATEGORY BY USERS

| Application Type | Responds | Rate |
|---|---|---|
| Social Networking (e.g., Telegram, WhatsApp) | 43 | 97.7% |
| Medical (e.g., Diabetes app) | 6 | 13.6% |
| Entertainment (e.g., YouTube) | 31 | 70.5% |
| Health and Fitness (e.g., Google Fit) | 16 | 36.4% |
| Games (Any games) | 12 | 27.3% |

### B. Study Objectives

Most healthcare professionals struggle with keeping track of their patients after their hospital discharge. This is often handled with regular check-ups with their patients which is a cumbersome for both doctors and patients. More specifically, elderly patient who had hospital discharge living in rural areas or places where commuting is an issue find it really challenging to keep track of their health condition with their doctor. Therefore, a chatbot system integrated with a telemedicine system is an ideal solution for this issue. Telemedicine should tailor the health service to patients needs offering valuable content to enhance their health condition. Telemedicine systems need to provide patients with assistance, feedback and encouragement to alleviate any feelings of detachment during the patient support. We discuss chatbot role in telemedicine application to support elderly patients living in rural areas after their hospital discharge. The discussed system encompasses technical, design, behavioural, and linguistic aspects in its general architecture. Patients can access through the chatbot system various health services and support, including doctor appointment, medication adherence, diet and physical activity promotion, and health coaching all after the hospital discharge. The bot will also check aspects related to user emotional state and sentiment at a certain time. The patient can chat with the bot about their feeling and condition after their medication, based on that the bot will provide them with personal recommendations. In case the suggestion is not effective, the bot can suggest direct contact from the doctor. To sum up the objectives, we provide the following list of objectives of our system:

- To provide access to timely and efficient healthcare to elderly living in rural & remote areas.
- To reduce the cost associated with hospital visit and doctor check.
- To improve diagnosis and treatment facilities in rural areas.
- To mitigate the obstacles due to geographical isolation and age impairments.
- To provide continuous medical care to the patients in rural/remote areas after hospital discharge and facilitate the work of healthcare professionals.

### C. Chatbot Architecture for Telemedicine Domain

*1) Chatbot Characteristics:* Designing habits requires addressing specific domains with functionalities and features to fulfill certain action. Many domain oriented chatbots exist which are able to understand and respond to specific knowledge domain with multi-purpose initiatives and human-like behaviour. The challenge with chatbot function is understanding the conversation context and natural language analysis so to understand user request. Among the important characteristics of a telemedicine chatbot are its simplicity, adaptiveness, constant presence and motivation. The simplicity is for the interaction, since user don't have to learn the meaning of functionalities of menus. Besides it opens the system to elderly with technical impairment. Adaptiveness is related to its dialoguing nature and applied functionalities. The bot is constantly available and will always be there to provide support to patients. Chatbots give patients emotional support motivating them with friendly feedback on their requests. This could be a combination of the chatbot + healthcare providers support and sometimes, patients won't even know that a chatbot was involved, which would likely result in a higher acceptance of the chatbot being human.

*2) Conversational User Interface Design:* This involves various design techniques applied in Conversational UI design, which is different than the common GUI design patterns. Among the general considerations in CUI design are the personality of the chatbot, notification for user engagement, flexibility in response, text vs custom keyboard, rich NLP, empathy and emotional state, keeping conversation short, and focusing on decreased user boredom [10]. These are some of the elements to consider when dealing with domain specific CUI design patterns.

*3) Behavioural Intervention Technology Model:* Chatbots should act in a certain way to respond to crucial emotional states of patients. For example, the bot should provide emotional feedback in case user emotion is detected. If the patient says "I felt bad last night after dinner", then the bot knows this is a negative emotion and its about patients condition. When emotional content is detected, the bot's feelings toward that person is adjusted. The chatbot should handle an optimist (broader and higher positive reactions), pessimist (broader and higher negative reactions), or moody feelings (high positive and negative reactions).

The chatbot should act in a certain way to respond to crucial emotional states by the patients. For example, making the bot provide proper feedback in case user is feeling bad. Within the telemedicine field there a specifically designed technological model to deliver behaviour change support and improve user's health. For our study we follow the Behavioural Intervention Technology Model (BITs) [11] to cover the theoretical and behavioural gaps in our system.

Human support has been integrated into BITs in different ways (text messaging, email, phone calls provided by supporters with varying expertise including therapists, nurses and trainees ). This, however, requires developed models for providing the necessary assistance. The Efficiency Model of Support [12] has been proposed as a framework to guide the actions of supporters delivering BITs by helping them to effectively manage the interplay between information and intervention. Efficiency is defined as the ratio of the outcome of an intervention relative to the human resources required to deliver it, since each decision corresponds to supporting that intervention (what, when, how much, and who provides it) represents a trade-off between devoting additional resources and accruing additional benefits. According to the model, decisions should be based on consideration of why people may fail to benefit from BITs and five categories of possible failure points are proposed: usability, engagement, fit, knowledge and implementation.

In telemedicine interventions for patient support, the deployment of a chatbot may lower usability barriers for users, since conversational agents are considered among the most intuitive to use kinds of BITs, requiring less learning effort by the user. By helping users to unobtrusively keeping track of their overall health condition, the chatbot can be an ideal solution to help them acquire an accurate knowledge of health related guidelines and to turn this knowledge into healthy habits and practices of daily life.

*4) Language Understanding:* This is strongly connected with the natural language understanding semantics used in the chatbot to enrich the natural communication with patients. As chatbots recognise and respond properly to a user's text messages. Due to the complexity of NL processing, the user is constrained to a set of available commands. Providing a solid language and context understanding requires having a strong knowledge-base. This gives the bot access to knowledge of answers to questions about various health issues. It integrates with the semantic web, querying realtime data from major datasources on the web. The bot should also have a story of events to make patient interaction easier, since we're dealing with elderly. Finally, the bot should compound sentences and sentiment detection capabilities. This generally helps the flow of conversation and detect patients emotional states.

*5) The Artificial Intelligence Capabilities:* To provide the necessary intelligence for the bot we should consider behaviour type and the bot domain. For example, if the bot is intended for supporting patients and providing them with clinically valid information, then we need to have access to medical data, building empathy with patients, and provide them with wise answers and intelligent conversation. Other features include syntax analysis, in case of typo occurrence. Syntax analysis provides a possibility to identify the structure and meaning of the text with the help of machine learning models in REST API, identify parts of speech, extract necessary information from the sentences and create dependency threads for each sentence. It can be used to extract information about people. Syntax analysis also allows understanding sentiment about a certain product, feature and parse intents from chatbot users. Dealing with patients health means considering their sentiment and experience with the disease after hospital discharge. Therefore, understanding how they feel at a certain stage is important. We need to perform sentiment analysis to understand the sentiment expressed in a block of text: positive, negative or neutral.

To achieve the above objectives we will integrate REST API services into the chatbot system. In this way, we will utilise smart solutions to empower the chatbot intelligence. We will use the DialogFlow[1] to enrich the bot with NLP and machine learning capabilities. This platform provides various advanced features to the bot, such as geolocation and wisdom in conversation. The API provides support for key phrases patterns of words that the Bot listens to. This will include understanding context like, "are you", "will you" and "can you". To enrich the medical knowledge for the bot and provide it with information about symptoms and medications, we will use ApiMedic [2], which offers a medical symptom checker primarily for patients. Based on the entered symptoms it tells what possible diseases there could be. It directs patients to more medical information. Finally, to provide the bot with sentiment analysis power, we will use the MonkeyLearn[3] which is a highly scalable machine learning API to automate text classification, sentiment analysis, topic and language detection. To provide a user interface for the patient to communicate with the chatbot, we use the Telegram cloud-based instant messaging service API. This API provides encryption, and advanced CUI functionalities. For example, with the current keyboard functionality it provides, Telegram have made it increasingly easy to use their chatbots, which is an advantage for the elderly who often have issues with technologies, such as mobile applications.

### D. Telemedicine Chatbot Architecture

The chatbot developed can function as healthcare assistant. It should act doctor-like to provide responds to patients questions about their health issues. It could also act as companions for the elderly, home residence patients. To be more effective the chatbot systems would be extended to gather and store multi-modal data about the patient and the environment to better reason patients condition and needs. In Figure-1 we provide a high-level architecture of the telemedicine chatbot system. This architecture includes the patient-healthcare provider interaction, the bot communication with various external REST APIs and the information it provides to the telemedicine system.

## IV. SYSTEM FEATURES

The presented telemedicine chatbot system integrates emotions, knowledge base, sentence structure, unmatched pattern-matching capabilities empowered by REST APIs and machine learning components. The AI Engine is the core of the chatbot. It uses Natural Language Processing (NLP), where sentences are parsed and broken down to reveal the structure of the sentence and information about individual words and their relation to other words in the sentence. Responses are matched against both specific and broad categories of statements, and then the response is constructed using both the bot's own original words and a wealth of information available from the API services. The bot has long-term memory and will remember things about people, it understands the difference between a question and description of a symptoms and will react accordingly.

---

[1] https://docs.api.ai
[2] http://apimedic.com/
[3] http://monkeylearn.com/

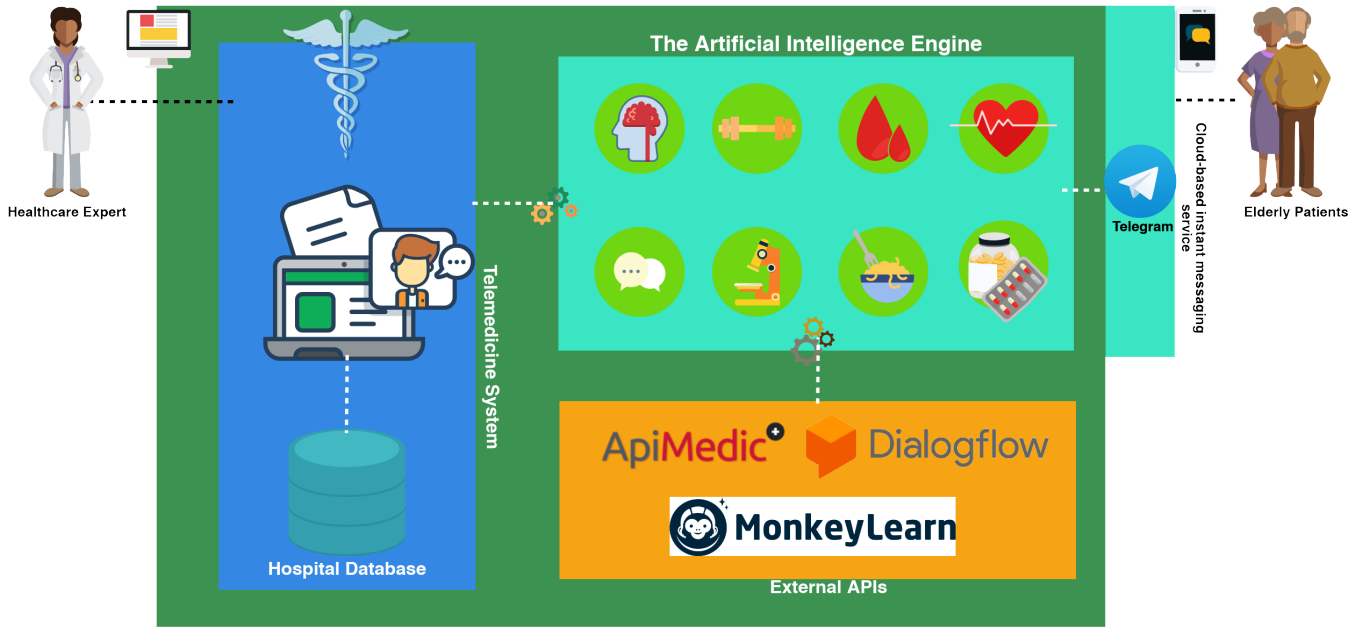

Fig. 1. The Telemedicine Chatbot Architecture

## V. SYSTEM DIAGRAM

The presented chatbot will gather information from patients related to their condition. The information must be elicited to reach a correct diagnosis. For example, the bot can ask the user a set of questions (such as, when did the pain start?, Where is the pain located?, What does the pain feel like?, Is the patient nauseous?). This interaction can be analysed and reported to the healthcare expert. The interaction will analyse the empathetic behaviour. Empathising with the patient is a key component of building rapport. Empathy lets the patient know the doctor understands their situation. Empathetic behaviour is also an indicator of the participant's emotional involvement in the interaction. Participants' empathetic actions (e.g., saying "I know it hurts," acknowledging the patient's fears, etc.) were tallied [7]. In Figure-2 we provide a block diagram representation of how the interaction happens between the patient and bot to conclude an activity. These processes guide the patient through a series of steps to achieve their daily goal. For instance, to exercise more or eat healthy, there has to be a finite set of steps a user has to fulfill to achieve it. The activity starts by **Access Bot** in which the conversation resides, before the conversation begins. The state changes to **Presentation**, where the bot provides necessary information about the service and possible functionalities offered by the bot. The bot provides the necessary information to the patient based on their request at the **Process** state. After that, at the **Options** state, the chatbot provides the activities for the patient based on their health condition (e.g., user's daily diet or physical activity plan, health and medication condition). The state then changes to **Satisfying** when the patient successfully fulfils the given activity and reaches the goal, and hence the state changes to **Conclude**, where the last goal-fulfilment task has been reached. The state then provides a feedback to the patient at the **Feedback** state, which in turn passes the information to the **Process** state to be available to the patient. The bot then closes the operation and changes state to **End**. However, when the state is **Unsatisfying**, then initially the system will try to provide other suggestions to the patient, if one will work then it will switch to **Satisfying** state and conclude. If changing suggestion won't work, then the system will request information from the **External Actor** state, which will involve a human actor in the process.

## VI. PATIENT-HEALTHCARE EXPERT SCENARIO

To better present the way a chatbot support may fit into a telemedicine domain, we provide a use case scenario of the patient-chatbot interaction aimed at providing health information and support for an elderly patient who recovered and discharged from hospital and lives in a rural area where direct healthcare service access is an issue. *"Sam is 66 years old, he lives in a small village outside the city. Sam is a diabetic patient who was discharged a month ago from the hospital. Currently he is living at home and having an appointment with*

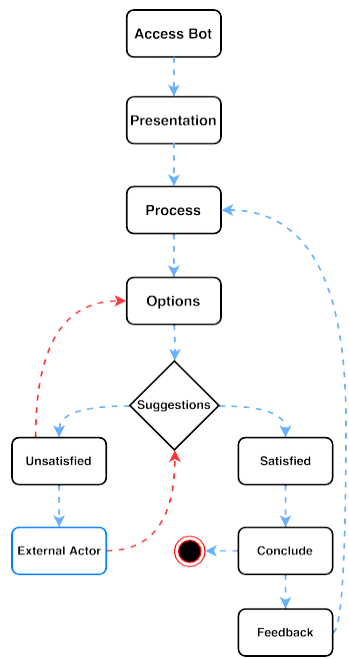

Fig. 2. The Patient-Chatbot Activity Diagram

*the doctor once in a while to check his condition. For Sam its a real struggle to arrange the appointment and commute to the clinic for the doctor visit. On the other hand, Dr. Silvia is an Endocrinologist and Dietitian who works part time at the hospital and at her clinic. Dr. Silvia has many diabetic patients and mostly are elderly people who find it hard to travel every time to meet her for 30 minutes. For Silvia sparing time for each patient is a burden and sometimes a physical appointment is unnecessary.*

*Dr. Silvia begins using the telemedicine-chatbot system and provides it to Sam. Now, Sam can directly chat with the bot about health tips and support. Moreover, Sam can get various recommendations related to diet, medication adherence, physical activity, and health condition. The bot can track Sam's condition and will notify the doctor in case of deterioration to intervene and provide direction support. This system made the life of Sam easier with respect to his disease experience. Sam can ask the bot directly about informations without bothering the doctor. After a long day, Dr. Silvia also uses the telemedicine system to talk to more patients who couldn't meet in person. As a doctor on a virtual platform she's been able to build amazing trust with many patients who keep coming back for her."*

## VII. DISCUSSION

There was no study, to our knowledge, that adapted conversational AI to investigate chatbots in telemedicine intended for elderly discharged from hospital and living in rural areas. We believe there is a room for improvement with respect to bot effectiveness and continues care. With this work, we discussed a telemedicine system that combines chatbot and advanced feature integration (e.g., machine learning models) to collect and understand various features about the patients. In this way, the bot can understand how the user feels with their health condition and trigger the right action that can better serve user preferences. Moreover, the bot triggers and notifies a healthcare expert (e.g., nutritionist) to intervene whenever relevant.

Chatbots can offer a lot of benefits in the telemedicine domain both for healthcare providers and patients. For example, using chatbots eases patients data logging, and patients might feel more human-like characteristics. However, bots can't provide emotional support, to some extend, to patients which is essential in healthcare. Therefore, we believe including a human expert in the loop is essential at least at some stages throughout patient's journey. Conversational interfaces are effective in making health easier to understand for those unfamiliar with it in health areas (e.g., physical activity promotion, hospital discharge instruction, explanation of medical documents, and family health history-tracking). Bots are effective in providing nutrition education and primary care services can harness the simplicity and age friendliness offered by bot in various health interventions. For example, unlike other mHealth apps, chatbots are age-friendly and can be adapted by age groups with technology barriers.

Health sector is where the conversational agents powered by AI and real human support will play a vital role. Another growth area will be predictive analytics, once the storage and retrieval of all this data is mastered. Benefits should be derived in mundane areas such as patient care. It is important to note that the improvements in healthcare are not limited to the first world, in fact the real changes will be seen in the treatment of patients in areas of the globe that rarely see a health professional. The UN's Sustainable Development Goals require that healthcare be radically improved in the most underserved areas [15].

## VIII. CONCLUSION

The quest of digital health for optimising patient engagement and the process of interacting and providing support to patients will benefit most from chatbot systems, especially after hospital discharge. This provides the possibility to streamline a patient-doctor communication over a chatbot powered telemedicine application. Using chatbots by elderly patients as alternatives will easily make them familiar with the technology, for its simplicity and efficiency in providing the support needed. In this paper, we discussed the application of chatbot systems in telemedicine application to support elderly patients living in rural areas after hospital discharge. The bot acts as a medical assistance to support patients in their health condition and accompany them in their healthcare journey. From the perspective of user interface design, the development of chatbots is less concerned with the visual, and focuses more on the narrative, conversational dimensions. Chatbots pose new challenges to the whole UX community, which requires their reorientation to the new platform. Finally, digital health will be rejuvenated by yet another digital transformation, this time spawned by the progress in conversational AI research. Future work will focus on implementing the chatbot platform and deploying it into a field testing with elderly patients.